\documentclass[12pt]{article}
\usepackage[cp1250]{inputenc}
\usepackage[verbose,a4paper,tmargin=0.5in,lmargin=1in,rmargin=1in]{geometry}
\usepackage[pdftex]{graphicx}
\usepackage{amsfonts}
\usepackage{indentfirst}
\usepackage{latexsym}
\usepackage{amsmath}
\usepackage{amssymb}
\usepackage{psfrag}
\usepackage{eufrak}
\usepackage[mathscr]{eucal} 

\title{Proper conformal symmetries in SD Einstein spaces.}

\author{$\textrm{Adam Chudecki}^{*}$
\ \ and \ \  $\textrm{Micha{\l}  Dobrski}^{*}$}

\begin{document}

\maketitle

$*$ Center of Mathematics and Physics, Technical University of Łódź, 
\newline
$\ \ \ \ \ $ Al. Politechniki 11, 90-924 Łódź, Poland, 
\newline
$\ \ \ \ \ $ adam.chudecki@p.lodz.pl, michal.dobrski@p.lodz.pl
\newline
\newline
\newline
\textbf{Abstract}. 

Proper conformal symmetries in self-dual (SD) Einstein spaces are considered. It is shown, that such symmetries are admitted only by the Einstein spaces of the type $[\textrm{N},-] \otimes [\textrm{N},-]$. Spaces of the type $[\textrm{N}] \otimes [-]$ are considered in details. Existence of the proper conformal Killing vector implies existence of the isometric, covariantly constant and null Killing vector. It is shown, that there are two classes of $[\textrm{N}] \otimes [-]$-metrics admitting proper conformal symmetry. They can be distinguished by analysis of the associated anti-self-dual (ASD) null strings. Both classes are analyzed in details. The problem is reduced to single linear PDE. Some general and special solutions of this PDE are presented.

% #############################################################################

\section{Introduction}

Presented work is devoted to some interesting aspects of complex and real relativity. Although the idea of employing complex numbers to the specetime geometry was considered by Einstein himself (who used the imaginary time coordinate in special theory of relativity), the essentially great interest in methods of complex analysis in relativity began in the sixties. The spinorial formalism introduced by R. Penrose \cite{intro_1} and the null tetrad formalism (G.C. Debney, R.P. Kerr, A. Schild \cite{intro_2}) were only the first step. In the seventies E.T. Newman showed \cite{intro_3, intro_4} that each asymptotically flat spacetime defines some 4-dimensional complex analytic differential manifold endowed with a holomorphic Riemannian metric. It has been proved that this metric satisfies the vacuum complex Einstein equations and the self-dual (SD) or anti-self-dual (ASD) part of its conformal curvature tensor (Weyl tensor) vanishes. Such spaces were called by Newman \textsl{the heavenly spaces} ($\mathcal{H}$-\textsl{spaces}).

$\mathcal{H}$-spaces have attracted a great deal of interest. In particular J.F. Plebański showed that vacuum complex Einstein equations for a $\mathcal{H}$-spaces can be reduced to a single second order nonlinear partial differential equation called \textsl{the heavenly equation} for one holomorphic function \cite{intro_5}. This equation, although strongly nonlinear, is integrable. The next step was the investigation of the structure of the $\mathcal{H}$-spaces with additional symmetry given by the isometric Killing vector \cite{biblio_10}. The classification of the $\mathcal{H}$-spaces admitting isometric Killing vector has been presented in \cite{intro_6}. It has been proved that there are five classes of heavenly spaces equipped with the isometric Killing vector. Three of them have been solved completely; in two other classes heavenly equation has been reduced to the equations well-known in mathematical physics. One of them is the Boyer - Finley - Plebański equation (BFP-equation) also known as the Toda field equation. Further works have explored the analogous classification in the case of homothetic Killing vectors \cite{intro_7, intro_8, biblio_9}. 

In 1976 J.F. Plebański and I. Robinson introduced the concept of \textsl{hyperheavenly spaces} ($\mathcal{HH}$-spaces) \cite{intro_9}. $\mathcal{HH}$-spaces appeared to be the natural generalization of the $\mathcal{H}$-spaces. Similarly, the existence of the isometric or homothetic Killing vector allowed to classify such spaces and find many explicit hyperheavenly metrics \cite{biblio_8, intro_10, intro_11}.

To the best our knowledge the problem of the so called \textsl{proper conformal symmetry} in $\mathcal{H}$-spaces has not been yet considered in details. Recall that a Killing vector $K_{b}$ is called \textsl{the proper conformal Killing vector} (pCKV) if $\nabla_{(a} K_{b)} = \chi \, g_{ab}$, $\nabla_{a} \chi \ne 0$. In Einstein spaces ($R_{ab}=0$, $R=-4 \Lambda$) such symmetries can appear only in the spaces of the types $[\textrm{N},-] \otimes [\textrm{N},-]$ \cite{biblio_4}. Real metric of the Lorentzian signature equipped with the pCKV is the special pp-wave metric (for detailed discussion of the pp-waves metrics with additional Killing symmetries see \cite{Maartens_Maharaj}). The generalization of this result for the complex type $[\textrm{N}] \otimes [\textrm{N}]$ (and, consequently, real type $[\textrm{N}] \otimes [\textrm{N}]$ with the signature $(++--)$) has been presented in \cite{biblio_3}. However, the heavenly case $[\textrm{N}] \otimes [-]$ (or equivalent type $[-] \otimes [\textrm{N}]$) has not been considered anywhere. The main aim of this paper is to fill this gap. 

Moreover, our considerations have geometrical meaning: there is a close relation between pCKV and the main geometrical objects essential in the theory of heavenly and hyperheavenly spaces, i.e. \textsl{the null strings}. Null string is a 2-dimensional, holomorphic surface, totally null and totally geodesic. $\mathcal{H}$ and $\mathcal{HH}$-spaces admit at least one congruence of such null strings. In Einstein spaces null Killing vectors generate SD and ASD null strings \cite{intro_7}. As will be shown the existence of the pCKV in Einstein spaces implies the existence of the second Killing vector which is isometric, covariantly constant and null. Therefore, pCKV generates the SD and ASD null strings.

Recently a great deal of interest has been devoted to the 4-dimensional Walker and Osserman spaces in neutral signature $(++--)$. Our considerations are generally complex, but they can be easily carried over to the case of the real spacetimes of the neutral signature. The presented metrics of the SD Einstein spaces with the pCKV appeared to be the most general metrics of the SD Einstein-Walker spaces with pCKV. 

Our paper is organized as follows. 

In section 2 we analyze the consequences of the existence of the pCKV. We point the relation between conformal factor $\chi$ and the SD and ASD null strings. It is shown, that a certain invariant of the pCKV is in a deep geometrical relation with the optical properties of congruence of ASD null strings. This relation allows to split the issue into two classes. Then we present the general results: the form of the pCKV, associated null isometric Killing vector and the form of the metric. In particular it is pointed out, that Einstein field equations reduce to the single, linear, second order PDE (\ref{rownanie_do_rozwiazania}). Some comments on the real 4-dimensional spaces admitting pCKV ends the section. 

Section 3 is devoted to the solutions of the equation (\ref{rownanie_do_rozwiazania}). This equation splits in two, essentially different equations, which generate two classes of the type $[\textrm{N}] \otimes [-]$ metrics with pCKV. Class I has been solved completely, but Class II, based on the Eqs. (\ref{rownanie_trudniejsze_xy}) appeared to be more complicated. As far as we were able to check, Eq. (\ref{rownanie_trudniejsze_xy}) is quite problematic. The reduction of this equation to the canonical form is not helpful. We discuss some methods of solving this equation based on the solution of simpler equation for the curvature coefficient $C^{(1)}$ (\ref{rownanie_na_krzywizne_xy}). The general solution (\ref{solutionnnn_for_the_ClassII}) is rather symbolical then explicit, but it allows to generate some special explicit metric (compare \ref{przyklad_metryki_klasy_II}). In section 4 we find some examples of the metrics of considered spaces. Concluding remarks ends the paper.

% #############################################################################

\setlength\arraycolsep{2pt}
\setcounter{equation}{0}

\section{Proper conformal symmetries.}

\subsection{Proper conformal Killing symmetries via geometry of null strings.}
\label{subsekcja_proper_conformal_Killing}

The Killing equations read
\begin{equation}
\label{podstawowe_nasze_rownania_Killinga}
\nabla_{(a} K_{b)} = \chi \, g_{ab}
\end{equation}
If $\chi=0$ then the Killing vector $K_{b}$ is called \textsl{isometric}, if $\chi=\textrm{const}$ then $K_{b}$ is \textrm{homothetic}, finally, if $\nabla_{a} \chi \ne 0$ then $K_{b}$ is called \textsl{proper conformal Killing vector}. [To be more precise, proper conformal Killing vector is called \textsl{special} if $\nabla_{b} \nabla_{a} \chi =0$ and \textsl{non-special} if $\nabla_{b} \nabla_{a} \chi \ne 0$. In Einstein spaces with $\Lambda=0$ there is always $\nabla_{b} \nabla_{a} \chi =0$ (compare (\ref{integrability_N})) so pCKV considered here are always \textsl{special}]. In what follows we consider proper conformal Killing vectors so we assume, that $\nabla_{a} \chi \ne 0$. The spinorial equivalents of the equations (\ref{podstawowe_nasze_rownania_Killinga}) read
\begin{equation}
\label{Killing_equations_spinorowo}
\nabla_{A}^{\ \; \dot{B}} K_{C}^{\ \; \dot{D}} = l_{AC} \in^{\dot{B}\dot{D}} + l^{\dot{B}\dot{D}} \in_{AC} - 2 \chi  \in_{AC} \in^{\dot{B}\dot{D}}
\end{equation}
with symmetric spinors
\begin{equation}
l_{AC} := \frac{1}{2} \nabla_{(A}^{\ \ \dot{N}} K_{C)\dot{N}} \ \ \ \ \ \ \ \ \ \ \ \ 
l^{\dot{B}\dot{D}} := \frac{1}{2} \nabla^{N(\dot{B}} K_{N}^{\ \; \dot{D})}
\end{equation}

In \cite{biblio_4} the integrability conditions of (\ref{Killing_equations_spinorowo}) have been studied. It is well known \cite{biblio_1} that proper conformal symmetries in Einstein spaces ($C_{AB\dot{C}\dot{D}}=0$, $R=-4 \Lambda$) with nonzero cosmological constant $\Lambda$ can appear only in the conformally flat spaces (de-Sitter spacetime). Assuming, that at least one of the Weyl curvature spinors is nonzero we must set $\Lambda =0$ (a nice and short proof of this fact can be found in \cite{biblio_2}). Moreover, from integrability conditions $C^{N}_{\ \; ABC} \nabla_{N}^{\ \; \dot{A}} \chi =0 =C^{\dot{N}}_{\ \; \dot{A}\dot{B}\dot{C}} \nabla^{A}_{\ \; \dot{N}} \chi$ we conclude, that $C_{ABCD}$ and $C_{\dot{A}\dot{B}\dot{C}\dot{D}}$ must be of the type $[\textrm{N}]$ or $[-]$, so proper conformal symmetries exist only in the spaces of the type $[\textrm{N},-] \otimes [\textrm{N},-]$. 

The case $[\textrm{N}] \otimes [\textrm{N}]$ has been completely solved in \cite{biblio_3}. It admits Lorentzian slice which is special pp-wave metric. Generally, we do not consider complex de Sitter spacetime here. The only cases, which remain to be considered are spaces of the type $[\textrm{N}] \otimes [-]$ or equivalent spaces of the type $[-] \times [\textrm{N}]$. Here we deal with the type $[\textrm{N}] \otimes [-]$, so $C_{\dot{A}\dot{B}\dot{C}\dot{D}}=0$. In what follows $C_{ABCD} \ne 0$. Integrability conditions of proper conformal Killing equations in SD Einstein spaces read
\begin{subequations}
\begin{eqnarray}
\label{integrability_L_undot}
&& \nabla_{R}^{\ \; \dot{A}} l_{ST} + 2C^{N}_{\ RST} K_{N}^{\ \; \dot{A}}  + 2 \in_{R(S} \nabla_{T)}^{\ \ \dot{A}} \chi =0 
\\
\label{integrability_L_dot}
&& \nabla^{A}_{\ \dot{R}} l_{\dot{S}\dot{T}} 
 + 2 \in_{\dot{R} ( \dot{S}} \nabla^{A}_{\ \ \dot{T})} \chi=0
\\
\label{integrability_M_undot}
&& K_{N\dot{N}} \nabla^{N\dot{N}}C_{ABCD} + 4C^{N}_{\ \; (ABC} l_{D)N} - 4 \chi C_{ABCD} = 0
\\
\label{integrability_N}
&& \nabla_{A}^{\ \; \dot{A}} \nabla_{B}^{\ \; \dot{B}} \chi  =0
\\
\label{integrability_R_undot}
&& C^{N}_{\ \; ABC} \nabla_{N}^{\ \; \dot{A}} \chi =0
\end{eqnarray}
\end{subequations}
From (\ref{integrability_R_undot}) we find, that $\nabla_{N}^{\ \; \dot{A}} \chi$ for $\dot{A}=\dot{1}, \dot{2}$ is a quadruple undotted Penrose spinor. However, multiple Penrose spinors for the type [N] must be necessarily linearly dependent, so
\begin{equation}
\label{warunek_na_liniowa_zaleznosc_spinP}
\det \, (\nabla_{N}^{\ \; \dot{A}} \chi) = 0 \ \Longleftrightarrow \ \nabla^{N\dot{A}} \chi \cdot \nabla_{N\dot{A}} \chi =0
\end{equation}
From (\ref{warunek_na_liniowa_zaleznosc_spinP}) it follows, that vector $\widehat{K}^{A\dot{B}} := \nabla^{A\dot{B}} \chi $ is null. It can be always presented in the form
\begin{equation}
\label{postac_izometrii_generowanej_przez_conf}
\widehat{K}^{A\dot{B}} := \nabla^{A\dot{B}} \chi = \mu^{A}\nu^{\dot{B}}
\end{equation}
where $\mu^{A}$ and $\nu^{\dot{B}}$ are some nonzero spinors. From (\ref{integrability_N}) it follows that vector $\widehat{K}^{A\dot{B}}$ is covariantly constant. But it means, that $\widehat{K}^{A\dot{B}}$ is an isometric Killing vector. We conclude, that existence of proper conformal Killing vector $K^{A\dot{B}}$ implies the existence of the second Killing vector $\widehat{K}^{A\dot{B}}$ which is isometric, null and covariantly constant. Inserting (\ref{postac_izometrii_generowanej_przez_conf}) into (\ref{integrability_R_undot}) we find that $C_{ABCD} = C \, \mu_{A}\mu_{B}\mu_{C}\mu_{D}$, $C\ne0$, so $\mu_{A}$ is a quadruple Penrose spinor. Inserting (\ref{postac_izometrii_generowanej_przez_conf}) into (\ref{integrability_N}) one gets the equations
\begin{subequations}
\begin{eqnarray}
\label{definicje_struny_podstawowej_SD}
&&\nabla_{A \dot{B}} \mu_{C} = Z_{A \dot{B}} \mu_{C} 
\\ 
\label{definicje_struny_podstawowej_ASD}
&&\nabla_{A \dot{B}} \nu_{\dot{C}} = - Z_{A \dot{B}} \nu_{\dot{C}}
\end{eqnarray}
\end{subequations}
where $Z_{A \dot{B}}$ are some spinors. From (\ref{definicje_struny_podstawowej_SD}) it follows that spinor $\mu_{A}$ satisfies the nonexpanding self-dual null string equation and from (\ref{definicje_struny_podstawowej_ASD}) it follows, that spinor $\nu_{\dot{A}}$ satisfies the nonexpanding anti-self-dual null string equation.

[Remark. Define the 2-dimensional, self-dual distribution 
\begin{equation}
\label{self_dual_distribution_D}
\mathcal{D}_{\mu^{A}} := \{ \mu_{A} \nu_{\dot{B}}  , \mu_{A} n_{\dot{B}} \} \ , \ \ \ \nu_{\dot{A}} n^{\dot{A}} \ne 0
\end{equation}
which is given by the Pfaff system 
\begin{equation}
\label{Pfaff_system_self_dual}
\mu_{A} \, g^{A \dot{B}} = 0 \ , \ \ \ 
(g^{A\dot{B}}) := \sqrt{2}
\left[\begin{array}{cc}
e^4 & e^2 \\
e^1 & -e^3
\end{array}\right] 
\end{equation}
where $(e^{1}, e^{2}, e^{3}, e^{4})$ is the null tetrad. Distribution $\mathcal{D}_{\mu^{A}}$ is integrable in the Frobenius sense iff the spinor $\mu_{A}$ satisfies the equation
\begin{equation}
\label{spinnorowe_rownanie_samodualnej_wstegi}
\mu^{A} \mu^{B} \nabla_{A \dot{B}} \mu_{B} = 0
\end{equation}
Then the integral manifolds of the distribution $\mathcal{D}_{\mu^{A}}$ are totally null and geodesic, self-dual holomorphic 2-surfaces (\textsl{self-dual null strings}) and they constitute the \textsl{congruence of the self-dual null strings}. Such congruence is called \textsl{nonexpanding}, if it is parallely propagated, which involves conditions stronger then the (\ref{spinnorowe_rownanie_samodualnej_wstegi}), i.e.
\begin{equation}
\mu^{B} \nabla_{A \dot{B}} \mu_{B} = 0
\end{equation}
Analogously one can define 2-dimensional, anti-self-dual distribution 
\begin{equation}
\dot{\mathcal{D}}_{\nu^{\dot{A}}} := \{ \mu_{A} \nu_{\dot{B}} , l_{A} \nu_{\dot{B}} \} \, \ \ \ \mu_{A} l^{A} \ne 0
\end{equation}
given by the Pfaff system
\begin{equation}
\nu_{\dot{B}} \, g^{A \dot{B}} = 0
\end{equation}
Distribution $\dot{\mathcal{D}}_{\nu^{\dot{A}}}$ is integrable in the Frobenius sense iff the spinor $\nu_{\dot{A}}$ satisfies the equation
\begin{equation}
\nu^{\dot{A}}\nu^{\dot{B}} \nabla_{A \dot{A}} \nu_{\dot{B}} = 0
\end{equation}
Integral manifolds of the distribution $\dot{\mathcal{D}}_{\nu^{\dot{A}}}$ are totally null and geodesic, anti-self-dual 2-surfaces (\textsl{anti-self-dual null strings}) and they constitute \textsl{congruence of anti-self-dual null strings}. If the congruence given by the distribution $\dot{\mathcal{D}}_{\nu^{\dot{A}}}$ is parallely propagated, i.e.
\begin{equation}
\nu^{\dot{B}} \nabla_{A \dot{A}} \nu_{\dot{B}} = 0
\end{equation}
then is called \textsl{nonexpanding}.]

Using (\ref{definicje_struny_podstawowej_SD}) in (\ref{integrability_M_undot}) we find the form of the spinor $l_{AB}$
\begin{equation}
\label{postac_spinora_l_AB}
l_{AB} = \mu_{(A} l_{B)} \ , \ \ \ \mu_{A}l^{A} \ne 0
\end{equation}
where $l_{A}$ is some spinor.

From the last integrability condition (\ref{integrability_L_dot}) we find that $l_{\dot{A}\dot{B}}$ is of the form
\begin{equation}
\label{postac_dotowanego_spinora_l}
l_{\dot{A}\dot{B}} = \eta_{(\dot{A}} \tau_{\dot{B})} \ , \ \ \ \eta_{\dot{A}} \tau^{\dot{A}} \ne 0
\end{equation}
and the following equations hold
\begin{subequations}
\begin{eqnarray}
\label{warunek_na_strune_ASD_eta}
&&\nabla_{A\dot{B}} \eta_{\dot{S}} = X_{A\dot{B}} \eta_{\dot{S}} + \in_{\dot{B}\dot{S}} \eta \mu_{A}
\\
\label{warunek_na_strune_ASD_tau}
&&\nabla_{A\dot{B}} \tau_{\dot{S}} = -X_{A\dot{B}} \tau_{\dot{S}} + \in_{\dot{B}\dot{S}} \tau \mu_{A}
\\ 
\label{warunek_na_ekspansje_eta_tau}
&& 2 \nu_{\dot{A}} + \eta \tau_{\dot{A}} + \tau \eta_{\dot{A}} =0
\end{eqnarray}
\end{subequations}
where $X_{A\dot{B}}$ is some spinor and $\eta$ and $\tau$ are some scalar functions. 

From (\ref{warunek_na_strune_ASD_eta}) and (\ref{warunek_na_strune_ASD_tau}) we easily find, that the distributions 
\begin{equation}
\dot{\mathcal{D}}_{\eta^{\dot{A}}} := \{ \mu_{A} \eta_{\dot{B}} , l_{A} \eta_{\dot{B}} \}
 \ , \ \ \ 
\dot{\mathcal{D}}_{\tau^{\dot{A}}} := \{ \mu_{A} \tau_{\dot{B}} , l_{A} \tau_{\dot{B}} \}
\end{equation}
are integrable in the Frobenius sense and their integrable manifolds constitute the congruences of the anti-self-dual null strings (with expansions given by the functions $\eta$ and $\tau$, respectively). Because spinor $\nu_{\dot{A}}$ must be nonzero, from (\ref{warunek_na_ekspansje_eta_tau}) we find, that at least one of the $\dot{\mathcal{D}}_{\eta^{\dot{A}}}$ or $\dot{\mathcal{D}}_{\tau^{\dot{A}}}$ must be expanding. 

It is worth to note, that for any pCKV the invariants $l_{AB}l^{AB}$ and $l_{\dot{A}\dot{B}} l^{\dot{A}\dot{B}}$ cannot be constant.

We end this general discussion by pointing out a condition very useful in further analysis. Using (\ref{postac_dotowanego_spinora_l}) and (\ref{warunek_na_ekspansje_eta_tau}) we find 
\begin{equation}
\label{warunekk_na_ekspanse_wsteeg}
l_{\dot{A}\dot{B}} \, \nu^{\dot{A}} \nu^{\dot{B}} = \frac{1}{2} \eta \tau \, l_{\dot{A}\dot{B}} l^{\dot{A}\dot{B}} 
\end{equation}
because $l_{\dot{A}\dot{B}} l^{\dot{A}\dot{B}} \ne 0$ from (\ref{warunekk_na_ekspanse_wsteeg}) it follows that
\begin{equation}
l_{\dot{A}\dot{B}} \, \nu^{\dot{A}} \nu^{\dot{B}} = 0 \ \ \ \Longleftrightarrow \ \ \ \eta \tau = 0
\end{equation}
so the invariant $l_{\dot{A}\dot{B}} \, \nu^{\dot{A}} \nu^{\dot{B}}$ can be used to distinguish two essentially different geometrical cases:
\begin{eqnarray}
\label{warunek_na_geometrie_dotowanych_strings}
&&l_{\dot{A}\dot{B}} \, \nu^{\dot{A}} \nu^{\dot{B}} \ne 0 \ \ \ \Longleftrightarrow \ \ \ \textrm{both } \dot{\mathcal{D}}_{\eta^{\dot{A}}} \textrm{ and } \dot{\mathcal{D}}_{\tau^{\dot{A}}} \textrm{ are expanding}
\\ \nonumber
&&l_{\dot{A}\dot{B}} \, \nu^{\dot{A}} \nu^{\dot{B}} = 0 \ \ \ \Longleftrightarrow \ \ \ \dot{\mathcal{D}}_{\eta^{\dot{A}}} \textrm{ or } \dot{\mathcal{D}}_{\tau^{\dot{A}}} \textrm{ is nonexpanding}
\end{eqnarray}

\subsection{SD Einstein spaces with proper conformal symmetry.}
\label{sekcja_Killing}

To solve Killing equations and find the form of the metric admitting proper conformal symmetry together with the forms of isometric and proper conformal Killing vectors, we use the results from \cite{biblio_9}. We omit long and tedious calculations and we give only final results. Every  type-[N] metric of the SD Einstein space admitting proper conformal Killing vector can be brought to the form
\begin{equation}
\label{finally_form_ofmetric}
\frac{1}{2} ds^{2} =  dy \underset{s}{\otimes} du -  dx \underset{s}{\otimes} dv - \frac{\partial \epsilon}{\partial v} \, du \underset{s}{\otimes} du
\end{equation}
where $(u,v,x,y,)$ are appropriate coordinates. This metric admits pCKV $K$ and isometric, null, covariantly constant Killing vector $\widehat{K}$
\begin{eqnarray}
&& \widehat{K} = \frac{\partial}{\partial y}
\\
\label{konforemny_wektor_Killinga}
&& K= \big( a_{0} v + u^{2} +n_{0} u + k_{0} \big) \frac{\partial}{\partial u} + \big( v(u-b_{0}-n_{0}) + r_{0} u +s_{0} \big) \frac{\partial}{\partial v}
\\ \nonumber
&& \ \ \ \ +\big( x(u+b_{0}+n_{0}) + a_{0} y \big) \frac{\partial}{\partial x} + \big( 2a_{0} \epsilon - n_{0} y + (v+r_{0} )x \big) \frac{\partial}{\partial y}
\end{eqnarray}
The conformal factor corresponding to $K$ reads $\chi=u$.
The function $\epsilon=\epsilon(u,v)$ has to satisfy the linear PDE 
\begin{eqnarray}
\label{rownanie_do_rozwiazania}
&&\big( a_{0}v +u^{2} + n_{0}u + k_{0}  \big) \frac{\partial^{2} \epsilon}{\partial v \partial u} 
+\big( v(u-b_{0}-n_{0}) +r_{0}u +s_{0}  \big) \frac{\partial^{2} \epsilon}{\partial v^{2}}
\\ \nonumber
&& \ \ \ \ \ \ \ \ \ \ \ \ \ \ \ \ \ \ \ \ \ 
+ (2u+2n_{0})  \frac{\partial \epsilon}{\partial v} -2a_{0} \frac{\partial \epsilon}{\partial u} =0
\end{eqnarray}
The only nonzero curvature coefficient and connection form in the null tetrad
\begin{equation}
\label{tetrada_zerowa_tutaj}
e^{1} = -du \ , \ \ \ e^{2} = -dy + \frac{\partial \epsilon}{\partial v} \, du \ , \ \ \ e^{3} = dv \ , \ \ \ e^{4} = -dx
\end{equation}
read
\begin{equation}
C^{(1)} = 2 \, \frac{\partial^{3} \epsilon}{\partial v^{3}}
\ , \ \ \ \Gamma_{31} = \frac{\partial^{2} \epsilon}{\partial v^{2}} \, du
\end{equation}
Note that the coordinates $(u,v,x,y)$ are related to the second Plebański coordinates $(q^{\dot{A}}, p^{\dot{B}})$ \cite{biblio_8} by
\begin{equation}
\label{przedefiniowanie_zmiennych}
q^{\dot{1}}=u \ , \ \ \ q^{\dot{2}} = v \ , \ \ \ p^{\dot{1}} = x \ , \ \ \ p^{\dot{2}} = y
\end{equation}
and the form of the key function which generates the metric (\ref{finally_form_ofmetric}) is
\begin{equation}
\Theta = \frac{1}{2} \frac{\partial \epsilon}{\partial v} x^{2} 
\end{equation}

[Remark. All SD Einstein spaces of the type [N] (without any additional Killing symmetry) has been found in \cite{biblio_10}. Using slightly different notation for the coordinates, namely $u=-p$, $v=-y$, $x=q$ and $y=-x$ we find, that the tetrad (\ref{tetrada_zerowa_tutaj}) takes exactly the same form, as the tetrad (5.20) in \cite{biblio_10} . Function $F$ from \cite{biblio_10} is equal $F=2 (\partial \epsilon / \partial q^{\dot{2}})$. The second possible type-[N] heaven (tetrad (5.24) from \cite{biblio_10}) is equipped with expanding congruence of self-dual null strings. However, proper conformal Killing symmetry implies the existence of nonexpanding congruence of self-dual null strings (compare (\ref{self_dual_distribution_D})). It means, that the type-[N] heavens given by the tetrad (5.24) from \cite{biblio_10} does not admit any proper conformal symmetry.]

Undotted spinorial basis in (\ref{finally_form_ofmetric}) is chosen in such a manner that $\mu_{1}=0, \mu_{2} \ne 0$. Remembering that $\chi=u$, we find from (\ref{postac_izometrii_generowanej_przez_conf}), that the dotted spinor $\nu^{\dot{A}}$ must satisfy $\nu_{\dot{1}} \ne 0$, $\nu_{\dot{2}} =0$. Then
\begin{equation}
l_{\dot{A}\dot{B}} \nu^{\dot{A}} \nu^{\dot{B}} = -2a_{0} \nu_{\dot{1}}^{2}
\end{equation}
It follows, that the constant $a_{0}$ has a deep geometrical meaning (compare (\ref{warunek_na_geometrie_dotowanych_strings})) and the metrics (\ref{finally_form_ofmetric}) split into two
\begin{eqnarray}
\nonumber
\textrm{Class I:} && a_{0} =0 \ \ \ \textrm{one of the distributions $\dot{\mathcal{D}}_{\eta^{\dot{A}}}$ or  $\dot{\mathcal{D}}_{\tau^{\dot{A}}}$ is nonexpanding}
\\
\nonumber
\textrm{Class II:} && a_{0} \ne 0 \ \ \ \textrm{both $\dot{\mathcal{D}}_{\eta^{\dot{A}}}$ and $\dot{\mathcal{D}}_{\tau^{\dot{A}}}$ are necessarily expanding}
\end{eqnarray}

\subsection{Comments on real type-[N] SD Einstein spaces admitting pCKV}

An important question is to find all real slices of the complex SD Einstein spaces admitting pCKV. Lorentzian slices are admitted only by the complex spaces of the same self-dual and anti-self-dual Petrov-Penrose type. On the other hand, real spaces of the Euclidean signature $(++++)$ do not admit nontrivial null vectors. Existence of the null vector $\widehat{K}^{A \dot{B}}$ (\ref{postac_izometrii_generowanej_przez_conf}) proves, that the metric (\ref{finally_form_ofmetric}) does not admit any Euclidean slices. The only possible real slices of the metric (\ref{finally_form_ofmetric}) should be the ones of neutral (ultrahyperbolic) signature $(++--)$. 

It can be easily proved, that in the real case spinor $\mu_{A}$ is real. It is guaranteed by the existence of the null vector $\widehat{K}^{A \dot{B}}$ (\ref{postac_izometrii_generowanej_przez_conf}). Indeed, in the real case of the signature $(++--)$ the vector $\widehat{K}^{A \dot{B}}$ satisfies
\begin{equation}
\label{warunek_na_rzeczywistosc}
\widehat{K}^{A \dot{B}} = \bar{\widehat{K}}^{A \dot{B}} 
\end{equation}
(where bar stands for the complex conjugation). From (\ref{warunek_na_rzeczywistosc}) we conclude, that
\begin{equation}
\bar{\mu}^{A} = \mu \, \mu^{A} \ , \ \ \ \bar{\nu}^{\dot{A}} = \frac{1}{\mu} \, \nu^{\dot{A}} \ , \ \ \ \mu \bar{\mu} = 1
\end{equation}
where $\mu$ is complex. However, spinors $\mu_{A}$ and $\nu_{\dot{A}}$ defined by (\ref{postac_izometrii_generowanej_przez_conf}) are not defined uniquely and they can be always re-defined according to
\begin{equation}
\mu^{A} \rightarrow e^{i \varphi} \mu^{A} \ , \ \ \ \nu^{\dot{A}} \rightarrow e^{-i \varphi} \nu^{\dot{A}}
\end{equation}
where $\varphi$ is some real function. Finally, using factor $\varphi$ one can always set $\mu=1$, making both $\mu^{A}$ and $\nu^{\dot{A}}$ real. Having $\mu^{A}$ real we can repeat the construction of the (real) Plebański tetrad based on the (real) self-dual null string defined by the spinor $\mu_{A}$. Summing up, all the results from subsection \ref{sekcja_Killing} can be used in the real case of the signature $(++--)$ without any loss of generality. It is enough to consider all the coordinates, functions and constants from subsection \ref{sekcja_Killing} as a real smooth objects, instead of the holomorphic ones.

% #############################################################################

\setcounter{equation}{0}
\section{Solutions of Eqs. (\ref{rownanie_do_rozwiazania})}

\subsection{Class I (case $a_{0}=0$)}

If $a_{0}=0$ then constants $n_{0}$ and $k_{0}$ become gauge-invariants, but the constant $r_{0}$ can be always gauged away. Moreover, since $a_{0}=0$ then the factor with $\epsilon$ in (\ref{konforemny_wektor_Killinga}) vanishes and to determine the metric we need only $\partial \epsilon / \partial v$. The equation (\ref{rownanie_do_rozwiazania}) takes the form
\begin{equation}
\big( u^{2} + n_{0}u + k_{0}  \big) \frac{\partial^{2} \epsilon}{\partial v \partial u} 
+\big( v(u-b_{0}-n_{0})  +s_{0}  \big) \frac{\partial^{2} \epsilon}{\partial v^{2}}
+ (2u+2n_{0})  \frac{\partial \epsilon}{\partial v}  =0
\end{equation}
with general solution
\begin{eqnarray}
\label{ogole_rozwiazanie_rroww_ClassI}
&&\frac{\partial \epsilon}{\partial v} = \exp \bigg( \! \! - \! \! \int \! \frac{2u+2n_{0}}{u^{2} +n_{0}u + k_{0}} \, du \bigg)  \, f(z)
\\
\nonumber
&&z:= v \, \alpha(u) - s_{0} \! \int \! \frac{\alpha(u) }{u^{2} +n_{0}u + k_{0}} \, du \ , \ \ \ \alpha(u) := \exp \bigg( \! \! - \! \! \int \frac{u-b_{0}-n_{0}}{u^{2} +n_{0}u + k_{0}} \, du \bigg) 
\end{eqnarray}
where function $f(z)$ is an arbitrary function of its variable.

\subsection{Class II (case $a_{0}\ne 0$)}
\label{subsekcja_Class_II}
%============KOMENDY======================
\newcommand{\vrx}{w} %Zmienna "x"
\newcommand{\vry}{z} %Zmienna "y"
\newcommand{\vrt}{s} %Zmienna typu "czas" w krzywych całkowych pola wekt.
\newcommand{\vrv}{\tilde{\vrx}} %Zmienna "v" w moich notatkach
\newcommand{\vru}{\tilde{\vrt}} %Zmienna "u" w moich notatkach
\newcommand{\prt}{g} %Oznaczenie na pierwiastek wielomianu 3 stopnia
\newcommand{\phif}[2]{\phi_{#1 , #2}}
\newcommand{\phifxy}[2]{\phif{#1}{#2}(\vrx,\vry)}
\newcommand{\opsmb}{R}
\newcommand{\varintgr}{\tau}
%============KOMENDY======================

If $a_{0} \ne 0$ then $n_{0}$ and $k_{0}$ can be gauged to zero. One of the invariant constants $a_{0}$, $r_{0}$ or $s_{0}$ can be brought to the arbitrary nonzero value, hence we fix $a_0=1$. 
%The equation (\ref{rownanie_do_rozwiazania}) takes the form
%\begin{equation}
%\label{rownanie_trudniejsze}
%\big( v +u^{2}  \big) \frac{\partial^{2} \epsilon}{\partial v \partial u} 
%+\big( v(u-b_{0}) +r_{0}u +s_{0}  \big) \frac{\partial^{2} \epsilon}{\partial v^{2}}
%+ 2u  \frac{\partial \epsilon}{\partial v} -2 \frac{\partial \epsilon}{\partial u} =0
%\end{equation}
%Differentiating (\ref{rownanie_trudniejsze}) twice by $\partial / \partial v$ we obtain the equation for the curvature coefficient $C^{(1)}$
%\begin{equation}
%\label{rownanie_na_krzywizne}
%\big( v+u^{2} \big) \, \frac{\partial C^{(1)}}{\partial u} + \big( v(u-b_{0}) +r_{0}u +s_{0}  \big) \, \frac{\partial C^{(1)}}{\partial v} + (4u-2b_{0}) \, C^{(1)} =0
%\end{equation}
It is also convenient to introduce variables 
\begin{equation}
\label{pomocnicza_zamiana_zmiennych}
\vrx := u \quad \quad \vry := v+u^{2}
\end{equation}
for which equation (\ref{rownanie_do_rozwiazania}) takes the following form
\begin{equation}
\label{rownanie_trudniejsze_xy}
\vry \, \frac{\partial^{2} \epsilon}{\partial \vry \partial \vrx} + \Big( \vry(3\vrx-b_{0}) -\vrx^{3} +b_{0}\vrx^{2} +r_{0}\vrx +s_{0}  \Big) \, \frac{\partial^{2} \epsilon}{\partial \vry^{2}} -2 \vrx \, \frac{\partial \epsilon}{\partial \vry} -2 \, \frac{\partial \epsilon}{\partial \vrx} = 0
\end{equation}
Differentiating it twice with respect to $\vry$ one obtains the equation for the curvature coefficient $C^{(1)}$
%\begin{equation}
%\label{rownanie_na_krzywizne_xy}
%\vry \frac{\partial C^{(1)}}{\partial \vrx} + \Big( \vry(3\vrx-b_{0}) -\vrx^{3} +b_{0}\vrx^{2} +r_{0}\vrx +s_{0}  \Big) \, \frac{\partial C^{(1)}}{\partial \vry} + (4\vrx-2b_{0}) \, C^{(1)} =0
%\end{equation}
\begin{multline}
\label{rownanie_na_krzywizne_xy}
\vry \frac{\partial C^{(1)}}{\partial \vrx} + \Big( \vry(3\vrx-\prt_1-\prt_2-\prt_3) -(\vrx-\prt_1)(\vrx-\prt_2)(\vrx-\prt_3)  \Big) \, \frac{\partial C^{(1)}}{\partial \vry}\\
+ \Big(4\vrx-2(\prt_1+\prt_2+\prt_3)\Big) \, C^{(1)} =0
\end{multline}
Here, we have rewritten the polynomial $\delta(\vrx)=-\vrx^{3} +b_{0}\vrx^{2} +r_{0}\vrx +s_{0}$ using its roots $\delta(\vrx)=-(\vrx-\prt_1)(\vrx-\prt_2)(\vrx-\prt_3)$. Hence, $b_0=\prt_1+\prt_2+\prt_3$, $r_0=-\prt_1 \prt_2 -\prt_1 \prt_3 - \prt_2 \prt_3$ and $s_0=\prt_1 \prt_2 \prt_3$. %We arrive at the following form of the equation for $C^{(1)}$

Relation (\ref{rownanie_na_krzywizne_xy}) is a first order linear partial differential equation for a function of two variables, so its general solution is determined by two independent first integrals. As far as we were able to check, it is rather improbable to guess these integrals directly from Lagrange--Charpit system for (\ref{rownanie_na_krzywizne_xy}). However, if one writes down equations for integral curves of corresponding vector field
\begin{equation}
\begin{split}
C^{(1)\prime}(\vrt)&=-\big(4 \vrx(\vrt)-2(\prt_1+\prt_2+\prt_3)\big)C^{(1)}(\vrt)\\
\vrx'(\vrt)&=\vry(\vrt)\\
\vry'(\vrt)&= \vry(\vrt)\big(3\vrx(\vrt)-\prt_1-\prt_2-\prt_3\big) -\big(\vrx(\vrt)-\prt_1\big)\big(\vrx(\vrt)-\prt_2\big)\big(\vrx(\vrt)-\prt_3\big)  
\end{split}
\end{equation}
than it immediately follows that $\vrx(\vrt)$ is determined by 
\begin{equation}
\vrx''(\vrt)= \vrx'(\vrt)\big(3\vrx(\vrt)-\prt_1-\prt_2-\prt_3\big) -\big(\vrx(\vrt)-\prt_1\big)\big(\vrx(\vrt)-\prt_2\big)\big(\vrx(\vrt)-\prt_3\big)
\end{equation}
This second order ODE admits Lie point symmetry\footnote{We used \texttt{Maxima} computer algebra system with \texttt{symmgrp} package \cite{symmgrp_paper} for determining this symmetry.} generated by the vector field
\begin{equation}
\vrx \frac{\partial}{\partial \vrt}+(\vrx-\prt_1)(\vrx-\prt_2)(\vrx-\prt_3) \frac{\partial}{\partial \vrx}
\end{equation}
It leads to the reduction
\begin{equation}
\label{rownanie_zwyczajne_zredukowane}
\vrv \frac{d\vrv}{d\vru}+(\vrv-\prt_1)(\vrv-\prt_2)(\vrv-\prt_3)=0
\end{equation}
with 
\begin{equation}
\begin{split}
\vru&=\int\frac{\vrx\,d\vrx}{(\vrx-\prt_1)(\vrx-\prt_2)(\vrx-\prt_3)}-\vrt\\ \vrv&=\frac{(\vrx-\prt_1)(\vrx-\prt_2)(\vrx-\prt_3)}{\vrx'}-\vrx. 
\end{split}
\end{equation}
Omitting longish details let us just mention, that having solutions of (\ref{rownanie_zwyczajne_zredukowane}) we were able to compute $\vrx(\vrt)$ from definition of $\vrv$, and in turn to extract first integrals from explicit formulas for $C^{(1)}(\vrt)$, $\vrx(\vrt)$, $\vry(\vrt)$. Notice however, that both $\vru$ and solutions of (\ref{rownanie_zwyczajne_zredukowane}) depend on degeneracy of roots of $\delta(\vrx)$. Thus, finally we are faced with three possible cases.
\newline
\begin{description}
\item[No degeneracy $\prt_1 \neq \prt_2$, $\prt_1 \neq \prt_3$, $\prt_2 \neq \prt_3$.]
For this case  the general solution of (\ref{rownanie_na_krzywizne_xy}) is given by
\begin{multline}
\label{gensol0}
C^{(1)}(\vrx,\vry)=\phi_{\prt_1,\prt_3}(\vrx,\vry)^{-2+\frac{2 \prt_3}{\prt_1-\prt_2}} \phi_{\prt_2,\prt_3}(\vrx,\vry)^{-2-\frac{2 \prt_3}{\prt_1-\prt_2}} \\
\times f\Big(\phi_{\prt_1,\prt_2}(\vrx,\vry)^{\prt_1-\prt_2}\phi_{\prt_3,\prt_1}(\vrx,\vry)^{\prt_3-\prt_1}
   \phi_{\prt_2,\prt_3}(\vrx,\vry)^{\prt_2-\prt_3}\Big)
\end{multline}
Here $\prt_3$ appears as a seemingly distinguished root. One can transform above solution to the form which does not favor any root
\begin{multline}
\label{gensol}
C^{(1)}(\vrx,\vry)=
\phifxy{\prt_1}{\prt_2}^{-\frac{2}{3}
   \left(\frac{\prt_1}{\prt_3-\prt_2}+\frac{\prt_2}{\prt_3-\prt_1}+2\right)} \phifxy{\prt_1}{\prt_3}^{-\frac{2}{3}
   \left(\frac{\prt_1}{\prt_2-\prt_3}-\frac{\prt_3}{\prt_1-\prt_2}+2\right)} \\
\times \phifxy{\prt_2}{\prt_3}^{-\frac{2}{3}
   \left(\frac{\prt_2}{\prt_1-\prt_3}+\frac{\prt_3}{\prt_1-\prt_2}+2\right)}\\
\times \tilde{f} \Big(\phifxy{\prt_1}{\prt_2}^{\prt_1-\prt_2} \phifxy{\prt_3}{\prt_1}^{\prt_3-\prt_1} \phifxy{\prt_2}{\prt_3}^{\prt_2-\prt_3}\Big)
\end{multline}
\item[Single degeneracy $\prt_1=\prt_2\neq \prt_3$.]
The general solution reads
\begin{equation}
\label{onesol}
C^{(1)}(\vrx,\vry)=
\frac{e^{ {-\frac{2 \prt_3 (\vrx-\prt_3)}{\phifxy{\prt_1}{\prt_3}}}}
   f\left(\frac{\phifxy{\prt_1}{\prt_1} 
\exp\big(\frac{(\prt_1-\prt_3)
   (\vrx-\prt_3)}{\phifxy{\prt_1}{\prt_3}}\big)}{\phifxy{\prt_1}{\prt_3}}\right)}{\phifxy{\prt_1}{\prt_3}^4}
\end{equation}
%\begin{equation}
%C(x,y)=
%\frac{e^{-\frac{2 \prt_3 (x-\prt_3)}{(x-\prt_1) (x-\prt_3)-y}}
%   f\left(\frac{\left((x-\prt_1)^2-y\right) e^{\frac{(\prt_1-\prt_3)
%  (x-\prt_3)}{(x-\prt_1) (x-\prt_3)-y}}}{(x-\prt_1)
% (x-\prt_3)-y}\right)}{((x-\prt_1) (x-\prt_3)-y)^4}
%\end{equation}
\item[Double degeneracy $\prt_1=\prt_2=\prt_3$.]
For this simplest case, we have found the following general solution 
\begin{equation}
\label{twosol}
C^{(1)}(\vrx,\vry)=\frac{e^{-\frac{2 \prt_1 (\vrx-\prt_1)}{\phifxy{\prt_1}{\prt_1}}} f\left(\frac{\phifxy{\prt_1}{\prt_1}-\vry}{\phifxy{\prt_1}{\prt_1}^2}\right)}{\phifxy{\prt_1}{\prt_1}^4}
\end{equation}
%\begin{equation*}
%C(x,y)=\frac{e^{-\frac{2 \prt_1 (x-\prt_1)}{(x-\prt_1)^2-y}} f\left(\frac{(x-\prt_1)^2-2y}{\left((x-\prt_1)^2-y\right)^2}\right)}{\left((x-\prt_1)^2-y\right)^4}
%\end{equation*}
\end{description}
In all above formulas $f$ is arbitrary function of single complex variable and $\phifxy{a}{b}:=(\vrx-a)(\vrx-b)-\vry$.

To reconstruct $\epsilon$ form $C^{(1)}$ one may proceed as follows. Transition from (\ref{rownanie_trudniejsze_xy}) to (\ref{rownanie_na_krzywizne_xy}) can be written in shortened form as a relation
\begin{equation}
\frac{\partial^2}{\partial \vry^2} \opsmb_{\epsilon} \equiv \opsmb_{C^{(1)}} \frac{\partial^3}{\partial \vry^3}
\end{equation}
between differential operators
\begin{multline*}
\opsmb_{\epsilon}=\vry \frac{\partial^2}{\partial \vrx \partial \vry} + \Big(\vry (3\vrx-\prt_1-\prt_2-\prt_3) -(\vrx-\prt_1)(\vrx-\prt_2)(\vrx-\prt_3)\Big)\frac{\partial^2}{\partial \vry^2}\\
-2\vrx\frac{\partial}{\partial \vry}-2\frac{\partial}{\partial \vrx}
\end{multline*}
and
\begin{multline*}
\opsmb_{C^{(1)}}=
\vry \frac{\partial }{\partial \vrx} +\Big(\vry(3 \vrx - \prt_1 - \prt_2 - \prt_3)  - (\vrx - \prt_1)(\vrx-\prt_2) (\vrx - \prt_3)\Big) \frac{\partial }{\partial \vry} \\
+ \Big(4 \vrx - 2 (\prt_1  +\prt_2 + \prt_3)\Big)
\end{multline*}
Thus, each solution $C^{(1)}(\vrx,\vry)$ of (\ref{rownanie_na_krzywizne_xy}), yields
\begin{equation}
\label{epsilon_prerozwiazanie}
\epsilon(\vrx,\vry)=\beta_0(\vrx)+\beta_1(\vrx)(\vry-\vry_0)+\beta_2(\vrx)(\vry-\vry_0)^2+\frac{1}{2}\int_{\vry_0}^{\vry} d\varintgr \int_{\vry_0}^{\varintgr}d\varintgr' \int_{\vry_0}^{\varintgr'}d\varintgr'' C^{(1)}(\vrx,\varintgr'') 
\end{equation}
as a solution of
\begin{equation}
\label{rownanie_trudniejsze_dwa_razy_zrozniczkowane}
\frac{\partial^2}{\partial \vry^2} \opsmb_{\epsilon} \epsilon=0
\end{equation}
with arbitrary functions $\beta_0(\vrx)$, $\beta_1(\vrx)$ and $\beta_2(\vrx)$. However, the solutions of $\opsmb_{\epsilon} \epsilon=0$ are of primary interest. Let us integrate equation (\ref{rownanie_trudniejsze_dwa_razy_zrozniczkowane}) twice from $\vry_0$ to $\vry$. %Using relation 
%\begin{equation*}
%\int_{\vry_0}^\vry d\varintgr_1 \int_{\vry_0}^{\varintgr_1} d\varintgr_2 \frac{\partial^2 f_{1}(\varintgr_2,w)}{\partial \varintgr_2^2}=f_{1}(\vry,w)-f_{1}(\vry_0,w)-(\vry-\vry_0)\frac{\partial f_{1}}{\partial \vry}(\vry_0,w)
%\end{equation*}
We immediately arrive at
\begin{equation*}
(\opsmb_{\epsilon}\epsilon)(\vrx,\vry)-(\opsmb_{\epsilon}\epsilon)(\vrx,\vry_0)-(\vry-\vry_0)\frac{\partial (\opsmb_{\epsilon}\epsilon)}{\partial \vry}(\vrx,\vry_0 )=0
\end{equation*}
holding true for arbitrary $\epsilon(\vrx,\vry)$ given by (\ref{epsilon_prerozwiazanie}). One constrains $\epsilon(\vrx,\vry)$ to satisfy $\opsmb_{\epsilon}\epsilon=0$ by requiring
\begin{equation}
\label{rownania_na_a_ogolne}
(\opsmb_{\epsilon}\epsilon)(\vrx,\vry_0)=0
\quad \text{and}\quad
\frac{\partial (\opsmb_{\epsilon}\epsilon)}{\partial \vry}(\vrx,\vry_0)=0
\end{equation}
After inserting (\ref{epsilon_prerozwiazanie}) into (\ref{rownania_na_a_ogolne}) the system of two ODEs for three functions emerges
%\begin{align}
%\nonumber
%\Big( (\vrx-\prt_1)(\vrx-\prt_2)(\vrx-\prt_3) + (\prt_1+\prt_2+\prt_3-3\vrx)\vry_0\Big)\beta_2(\vrx)& \\
%\label{rownania_na_a_sz1}
%+\vrx \beta_1(\vrx)+\beta_0'(\vrx)-\frac{1}{2}\vry_0 \beta_1'(\vrx)& =0\\
%\nonumber
%\frac{1}{2}\Big( (\vrx-\prt_1)(\vrx-\prt_2)(\vrx-\prt_3) + (\prt_1+\prt_2+\prt_3-3\vrx)\vry_0\Big) C^{(1)}(\vrx,\vry_0)& \\
%\label{rownania_na_a_sz2}
%+2(\prt_1+\prt_2+\prt_3-\vrx)\beta_2(\vrx) +\beta_1'(\vrx)-2\vry_0 \beta_2'(\vrx)& =0
%\end{align}
%where $\beta_{0}' := (d \beta_{0} / d w)$, etc. 
leading to straightforward but a bit cumbersome solutions for $\beta_0(\vrx)$, $\beta_1(\vrx)$, $\beta_2(\vrx)$. 
%It can be immediately observed that this system can be easily solved by taking $\beta_2(\vrx)$ as an a;rbitrary function, calculating $\beta_1(\vrx)$ from (\ref{rownania_na_a_sz2}) and then $\beta_0 (\vrx)$ from (\ref{rownania_na_a_sz1}). 
However, if we are allowed to take $\vry_0=0$ than aforementioned system of ODEs can be simplified to
\begin{align}
\label{rownania_na_a_2_sz1}
(\vrx-\prt_1)(\vrx-\prt_2)(\vrx-\prt_3)\beta_2(\vrx)+\vrx \beta_1(\vrx)+\beta_0'(\vrx)& =0\\
\label{rownania_na_a_2_sz2}
\frac{1}{2}(\vrx-\prt_1)(\vrx-\prt_2)(\vrx-\prt_3) C^{(1)}(\vrx,0)+2(\prt_1+\prt_2+\prt_3-\vrx)\beta_2(\vrx) +\beta_1'(\vrx)& =0
\end{align}
(here $\beta_0' = (d \beta_0 / d \vrx)$, etc.).
Taking as arbitrary function $\beta_1(\vrx)=\kappa(\vrx)$ we compute algebraically $\beta_2(\vrx)$ from (\ref{rownania_na_a_2_sz2}), and then integrate (\ref{rownania_na_a_2_sz1}) to get $\beta_0$. Consequently the following final formula for solutions of (\ref{rownanie_trudniejsze_xy}) can be given
\begin{eqnarray}
\label{solutionnnn_for_the_ClassII}
&&\epsilon=
-\frac{\kappa'(\vrx)+\frac{1}{2}(\vrx-\prt_1) (\vrx-\prt_2) (\vrx-\prt_3) C^{(1)}(\vrx,0)}{2(\prt_1+\prt_2+\prt_3-\vrx)}\vry^2+\kappa(\vrx)\vry+
\epsilon_0
\\ \nonumber
&&-\int_{0}^{\vrx} \bigg( \tilde{\varintgr} \kappa(\tilde{\varintgr}) -\frac{\kappa'(\tilde{\varintgr})+\frac{1}{2}(\tilde{\varintgr}-\prt_1) (\tilde{\varintgr}-\prt_2) (\tilde{\varintgr}-\prt_3) C^{(1)}(\tilde{\varintgr},0)}{2(\prt_1+\prt_2+\prt_3-\tilde{\varintgr})} (\tilde{\varintgr}-\prt_1) (\tilde{\varintgr}-\prt_2)(\tilde{\varintgr}-\prt_3)\, \bigg) d\tilde{\varintgr} 
\\ \nonumber
 && \ \ \ \ \ \ \ \ \ \ \ \ 
+ \frac{1}{2}\int_{0}^{\vry} d\varintgr \int_{0}^{\varintgr}d\varintgr' \int_{0}^{\varintgr'}d\varintgr'' C^{(1)}(\vrx,\varintgr'') 
\end{eqnarray}
with $C^{(1)}$ as a solution of (\ref{rownanie_na_krzywizne_xy}), arbitrary function $\kappa(\vrx)$, and integration constant $\epsilon_0$.

\setcounter{equation}{0}

\section{Examples}

\subsection{Class I}

In the Class I case we were able to obtain the general solution. Putting (\ref{ogole_rozwiazanie_rroww_ClassI}) into (\ref{finally_form_ofmetric}) we find the form of the metric
\begin{equation}
\label{metryka_classI_powstawieniu}
\frac{1}{2} ds^{2} =  dy \underset{s}{\otimes} du -  \frac{1}{\alpha} \, dx \underset{s}{\otimes} dz - b(u,z) \,  dx \underset{s}{\otimes} du -m(u) f(z) \, du \underset{s}{\otimes} du
\end{equation}
where $\alpha(u)$ is given by (\ref{ogole_rozwiazanie_rroww_ClassI}), $f(z)$ is an arbitrary function and we used the abbreviations
\begin{eqnarray}
&& b(u,z) := \frac{1}{\alpha} \, \frac{u-b_{0}-n_{0}}{u^{2}+n_{0}u+k_{0}} \bigg( z - s_{0} \! \! \int \! \frac{\alpha }{(u-b_{0}-n_{0})^{2}} \, du \bigg)
\\ \nonumber
&& m(u) := \exp \bigg(\! \! - \! \! \int \! \frac{2u + 2n_{0}}{u^{2} +n_{0}u+k_{0}} \, du \bigg)
\end{eqnarray}
The nonzero curvature coefficient is given by
\begin{equation}
C^{(1)} = 2 \exp \bigg(\! \! - \! \! \int \! \frac{4u - 2b_{0}}{u^{2} +n_{0}u+k_{0}} \, du \bigg) \frac{d^{2} \! f}{dz^{2}}
\end{equation}

\subsection{Case with the second isometric Killing vector}

In the next example we assume the existence of the second isometric Killing vector, namely
\begin{equation}
\tilde{K} = \frac{\partial}{\partial q^{\dot{1}}} = \frac{\partial }{\partial u}
\end{equation}
It gives strong restriction on function $\epsilon$
\begin{equation}
\label{warunek_na_epsilon_przy_drugim_Kil}
\frac{\partial^{2} \epsilon}{\partial u \partial v} = 0
\end{equation}
With the condition (\ref{warunek_na_epsilon_przy_drugim_Kil}) the equation (\ref{rownanie_do_rozwiazania}) can be easily solved and the solutions read
\begin{subequations}
\begin{eqnarray}
\textrm{Class I:} && \epsilon = a(u) + \frac{\epsilon_{0}}{v} \ , \ \ \ a_{0}=s_{0} = r_{0} =0 \ , \ \ \ b_{0}=-2n_{0}
\\
\textrm{Class II:} && \epsilon = \frac{1}{2} A_{0} \, u^{2} + B_{0} + \frac{\epsilon_{0}}{v+r_{0}} + A_{0} \, v
 \\ \nonumber
&& a_{0}=1 \ , \ \ \ s_{0}=b_{0}=n_{0}=k_{0}=0
\end{eqnarray}
\end{subequations}
where $a=a(u)$ is an arbitrary function and $A_{0}$, $B_{0}$ and $\epsilon_{0} \ne 0$ are constants. The corresponding metrics and the curvature take the form
\begin{subequations}
\begin{eqnarray}
\label{metryka_trzy_Killingi_1}
\textrm{Class I:} && \frac{1}{2} ds^{2} =  dy \underset{s}{\otimes} du -  dx \underset{s}{\otimes} dv + \frac{\epsilon_{0}}{v^{2}} \, du \underset{s}{\otimes} du \ , \ \ \ C^{(1)} = -\frac{12 \epsilon_{0}}{v^{4}}
\\
\label{metryka_trzy_Killingi_2}
\textrm{Class II:} && \frac{1}{2} ds^{2} =  dy \underset{s}{\otimes} du -  dx \underset{s}{\otimes} dv + \bigg( \frac{\epsilon_{0}}{(v+r_{0})^{2}} - A_{0} \bigg) du \underset{s}{\otimes} du \ , 
\\
\nonumber
&& C^{(1)} = -\frac{12 \epsilon_{0}}{(v+r_{0})^{4}}
\end{eqnarray}
\end{subequations}

\subsection{Special solution of the Class II}

The general solution of the Class II is complicated. We only present some simple example belonging to the Class II. We assume the double degeneracy of the polynomial $\delta$ with $g_{1}=g_{2}=g_{3}=0$ what corresponds to the $b_{0}=r_{0}=s_{0}=0$. Moreover we assume, that the arbitrary function $f$ in (\ref{twosol}) has the form $f(\xi) = (1 / \xi^{2})$, what gives expression for the curvature coefficient $C^{(1)}$
\begin{equation}
C^{(1)} = \frac{1}{(w^{2}-2z)^{2}} \stackrel{(\ref{pomocnicza_zamiana_zmiennych})}{=} \frac{1}{(2v+u^{2})^{2}}
\end{equation}
Following the procedure described in subsection \ref{subsekcja_Class_II} we find 
\begin{equation}
\epsilon = - \frac{1}{16} \, (2v+u^{2}) \Big( \ln (2v+u^{2}) + \omega_{u} \Big) - \frac{1}{16} v^{2} \, ( u^{-2} \omega_{u} - u^{-3} \omega) - \frac{1}{16} \, u^{2} + \epsilon_{0}
\end{equation}
where $\omega=\omega(u)$ is an arbitrary function and the $\epsilon_{0}$ is arbitrary constant. After simple calculations we obtain the form of the metric
\begin{equation}
\label{przyklad_metryki_klasy_II}
\frac{1}{2} ds^{2} =  dy \underset{s}{\otimes} du -  dx \underset{s}{\otimes} dv + \frac{1}{8} \Big(  \ln (2v+u^{2}) + \omega_{u} +v \, (u^{-2} \omega_{u} - u^{-3} \omega) +1 \Big) du \underset{s}{\otimes} du
\end{equation}

% #############################################################################

\section{Concluding remarks}

In this paper we investigated the problem of the proper conformal Killing symmetries is SD Einstein spaces of the type $[\textrm{N}] \otimes [-]$. We were able to find the forms of the proper conformal Killing vector (pCKV) and the generated by pCKV second Killing vector, covariantly constant and null. With this two Killing symmetries the problem has been reduced to the single, second order, linear PDE  (\ref{rownanie_do_rozwiazania}) for one function $\epsilon$, which completely determines the metric. 

One of us (AC) considered earlier \cite{biblio_3} algebraically less degenerated solution of the type $[\textrm{N}] \otimes [\textrm{N}]$, equipped with the pCKV. This case has been solved completely. It was obvious, that at least one class of the type $[\textrm{N}] \otimes [-]$ metrics with pCKV can be obtained from the metrics of the type $[\textrm{N}] \otimes [\textrm{N}]$ by simple algebraic reduction of the type. The question arose: are there any other classes of proper conformal Killing symmetries in type $[\textrm{N}] \otimes [-]$ which cannot be obtained from the type $[\textrm{N}] \otimes [\textrm{N}]$?

The answer for this question appeared to be positive. 
The final equation (\ref{rownanie_do_rozwiazania})  split into two classes. One of them (Class I) has been solved to the very end and it was this algebraic reduction of the type $[\textrm{N}] \otimes [\textrm{N}]$ metric. The second case appeared to be quite problematic, we achieved only partial success in solving it. Presented solution of the Class II must be considered rather like an algorithm how to generate the solutions from the curvature coefficient $C^{(1)}$. It was shown in section 5 how to use this procedure and obtain the explicit solutions of the Class II. 

It is worth to note, that the same problem can be attacked from the different side, namely: instead of the type $[\textrm{N}] \otimes [-]$ one can consider the type $[-] \otimes [\textrm{N}]$. Frankly, we followed that way and obtained nothing more, then the equation extremely similar to the equation (\ref{rownanie_do_rozwiazania}). The conclusion is, that using heavenly machinery in Plebański - Boyer - Finley coordinates nothing more can be done. However, we still believe, that form of the equation (\ref{rownanie_do_rozwiazania}) strongly depends on the coordinate frame. Maybe the Plebański - Boyer - Finley coordinates which we used in our work are simply not the optimal coordinates for the problem considered? Maybe the heavenly formalism in so-called first heavenly coordinates is more plausible? We are going to answer this question in future.

Finally we mention, that the metrics (\ref{metryka_classI_powstawieniu}), (\ref{metryka_trzy_Killingi_1}), (\ref{metryka_trzy_Killingi_2}) and (\ref{przyklad_metryki_klasy_II}) are interesting not only from the complex relativity point of view. Considering all the coordinates, functions and constant as a real smooth objects, we obtain the metrics of the neutral $(++--)$ signature. All of them are Einstein and equipped with both self-dual and anti-self-dual nonexpanding congruences of null strings. It proves, that they belong to the \textsl{two-sided Walker class} and \textsl{globally Osserman class} (compare \cite{concl_1}).

% #############################################################################


\begin{thebibliography}{99}



\bibitem{intro_1} Penrose R., \textsl{A spinor approach to general relativity}, Ann. Phys., \textbf{10}, 171 (1960)

\bibitem{intro_2}  Debney G.C., Kerr R.P. and Schild A., \textsl{Solutions of the Einstein and Einstein - Maxwell Equations}, J. Math. Phys., \textbf{10}, 1842 (1969)


\bibitem{intro_3}  Newman E.T., \textsl{The Bondi-Metzner-Sachs Group: Its Complexification and Some Related Curious Consequences}, Seventh International Conference on Gravitation and Relativity, Tel-Aviv (1974)


\bibitem{intro_4}  Newman E.T., \textsl{Heaven and its properties}, Riddle of Gravitation Symposium, Syracuse (1975)


\bibitem{intro_5}  Plebański J.F., \textsl{Some solutions of complex Einstein equations}, J. Math. Phys., \textbf{16}, 2395 (1975)


\bibitem{biblio_10} Finley J.D. III and Plebański J.F., \textsl{Further heavenly metrics and their symmetries}, J. Math. Phys., \textbf{17} 585 (1976)


\bibitem{intro_6} Finley J.D. III and Plebański J.F. \textsl{The clasification of all $\mathcal{H}$ spaces admitting a Killing vector} J. Math. Phys., \textbf{20}, 1938 (1979)


\bibitem{intro_7} Dunajski M. and West S., \textsl{Anti-self-dual conformal structures with null Killing vectors from projective structures}, Commun. Math. Phys., 272:85-118 (2007)


\bibitem{intro_8} Dunajski M., \textsl{An interpolating dispersionless integrable system}, Journal of Physics A: Mathematical and Theoretical, \textbf{41} 315202. (2008)


\bibitem{biblio_9} Chudecki A., \textsl{Classification of the Killing vectors in nonexpanding $\mathcal{HH}$-spaces with $\Lambda$}, Class. Quantum Grav., \textbf{29} 135010 (2012)


\bibitem{intro_9} Plebański J.F. and Robinson I., \textsl{Left - degenerate vacuum metrics}, Phys. Rev. Lett., \textbf{37} 493 (1976)



\bibitem{biblio_8} Plebański J.F. and Finley J.D. III, \textsl{Killing vectors in nonexpanding HH spaces}, J. Math. Phys., \textbf{19} 760 (1978)


\bibitem{intro_10} Sonnleitner S.A. and Finley J.D. III, \textsl{The form of Killing vectors in expanding $\mathcal{HH}$ spaces}, J. Math. Phys., \textbf{23(1)} 116 (1982)


\bibitem{intro_11} Chudecki A., \textsl{Homothetic Killing Vectors in Expanding $\mathcal{HH}$-Spaces with $\Lambda$}, International Journal of Geometric Methods in Modern Physics, Vol.10, No.1, 1250077 (2013)



\bibitem{biblio_4} Plebański J.F. and Hacyan S., \textsl{Some properties of Killing spinors}, J. Math. Phys., \textbf{17} 2204 (1976)


\bibitem{Maartens_Maharaj} Maartens R. and Maharaj S.D., \textsl{Conformal symmetries of pp-waves}, Class. Quant. Grav., \textbf{8}, 503-514 (1991)



\bibitem{biblio_3} Chudecki A., \textsl{Conformal Killing Vectors in Nonexpanding $\mathcal{HH}$-Spaces with $\Lambda$}, Class. Quant. Grav., \textbf{27} (205004) (2010)





\bibitem{biblio_1} Garfinkle D. and Tian Q., \textsl{Spacetimes with cosmological constant and a conformal Killing field have constant curvature}, Class. Quantum Grav., \textbf{4} 137 (1987)


\bibitem{biblio_2} Chudecki A. and Przanowski M., \textsl{Killing symmetries in $\mathcal{H}$-spaces with $\Lambda$}, J. Math. Phys., \textbf{54} 102503 (2013)




\bibitem{biblio_5} Finley J.D. III and Plebański J.F., \textsl{The intrinsic spinorial structure of hyperheavens}, J. Math. Phys., \textbf{17} 2207 (1976)


\bibitem{biblio_6} Boyer C.P., Finley J.D. III and Plebański J.F., \textsl{Complex general relativity, $\mathcal{H}$ and $\mathcal{HH}$ spaces - a survey to one approach}, General Relativity and Gravitation. Einstein Memorial Volume ed. A. Held (Plenum, New York) vol.2. pp. 241-281 (1980)



\bibitem{symmgrp_paper} Champagne B., Hereman W.  and  Winternitz P, \textsl{The computer calculation  of  Lie point symmetries of large systems of differential equations}, Comp. Phys. Comm., \textbf{66} 319 (1991)


\bibitem{concl_1} Chudecki A. and Przanowski M., \textsl{From hyperheavenly spaces to Walker and Osserman spaces: I}. Class. Quantum Grav., \textbf{25} 145010 (2008)



\end{thebibliography}
\end{document}